\documentclass[aps,prd, final]{revtex4}
\usepackage[latin1]{inputenc}
\usepackage{amsmath}
\usepackage{amsfonts}
\usepackage{amssymb}
\usepackage{amsthm}
\usepackage[dvips]{graphics}
\usepackage[pdftex]{graphicx}

\ifx\pdftexversion\undefined
\fi

\theoremstyle{remark}

\theoremstyle{definition}

\begin{document}

\title{Observable Effects of Scalar Fields and Varying Constants}
\author{John D. Barrow}
\email{J.D.Barrow@damtp.cam.ac.uk}
\affiliation{DAMTP, Centre for Mathematical Sciences, University of Cambridge,
Wilberforce Road, Cambridge CB3 0WA, UK}
\author{Douglas J. Shaw}
\email{D.Shaw@damtp.cam.ac.uk}
\affiliation{DAMTP, Centre for Mathematical Sciences, University of Cambridge,
Wilberforce Road, Cambridge CB3 0WA, UK}
\date{\today}

\begin{abstract}
We show by using the method of matched asymptotic expansions that a
sufficient condition can be derived which determines when a local experiment
will detect the cosmological variation of a scalar field which is driving
the spacetime variation of a supposed constant of Nature. We extend our
earlier analyses of this problem by including the possibility that the local
region is undergoing collapse inside a virialised structure, like a galaxy
or galaxy cluster. We show by direct calculation that the sufficient
condition is met to high precision in our own local region and we can
therefore legitimately use local observations to place constraints upon the
variation of "constants" of Nature on cosmological scales.

\end{abstract}

\maketitle

\section{Introduction}

The last few years have seen a resurgence of widespread interest in the
possibility that some or all of the fundamental `constants' of Nature might
be varying over cosmological timescales. To a large extent the revival in
this field has come about because of the need to explain and understand the
studies of relativistic fine structure in the absorption lines formed in
dust clouds around quasars carried out by Webb et al., \cite{webb}. Using a
new \textquotedblleft many-multiplet" method which exploits the information
in many wavelength separations of absorption lines with different
relativistic contributions to their fine structure considerable gains in
statistical significance were achieved. From a data set of 128 objects at
redshifts $0.5<z<3$, Webb et al. found their absorption spectra were
consistent with a relative in the value of the fine structure constant, $%
\alpha _{em}(z)$, between those redshifts and the present of $\Delta \alpha
_{em}/\alpha _{em}\equiv \alpha _{em}(z)-\alpha _{em}(0)/\alpha
_{em}(0)=-0.57\pm 0.10\times 10^{-5}$. In the seven years since their
results were first announced, extensive analysis has yet to identity any
systematic effect that could explain either its magnitude or sign. However,
a small study of 23 absorption systems in one hemisphere between $0.4\leq
z\leq 2.3$ by Chand et al., \cite{chand}, found a result consistent with no
variation: $\Delta \alpha _{em}/\alpha _{em}=-0.6\pm 0.6\times 10^{-6}$ but
this study uses only a simplified version of the many-multiplet method in
its analysis and concerns remain about calibrations and the noisiness of the
data fits. Later this year, a major effort to produce a very large new data
set should be reported and will clarify the status of these earlier
investigations. In the meantime there have been other astronomical checks on
the constancy of $\alpha _{em}$, using a variety of different techniques;
however, they have yet to reach the accuracy achieved using the
many-multiplet method. All found results \cite{jdbrs} that are consistent
with no variation but none could have seen the variation reported by Webb et
al. If $\alpha _{em}$ can, and does, vary in time then it would seem natural
to suspect that some other fundamental `constants' do so also. Recently,
using a study of the vibrational levels of $H_{2}$ in the absorption spectra
of quasars, Reinhold et al., \cite{reinhold}, reported a $3.5\sigma $
indiction of a variation in the electron-proton mass ratio, $\mu =m_{e}/m_{p}
$ over the last 12 Gyrs: $\Delta \mu /\mu =2.0\pm 0.6\times 10^{-5}$. This
result combines high-quality astronomical data with improvements in the
measurement of crucial laboratory wavelengths to new levels of precision. It
reports a variation at a level comparable to that claimed for $\alpha _{em}$
(which partly reflects state of the art precision in spectroscopic
measurements) but is theoretically much harder to understand. Variations in $%
\alpha _{em}$ and $\mu $ both lead to violations of the Weak Equivalence
Principle (WEP) and these are expected to be unacceptably large if $\Delta
\mu /\mu $ varies at the $10^{-5}$ level. Typically, the experimental
constraints on WEP violation from direct experimental studies of freefall
and the study of the relative motions of the Earth-Moon system lead to upper
bounds on the relative differential accelerations of different materials
which are $O(10^{-12})$. As explained in the detailed study by Barrow and
Magueijo \cite{bm}, the Webb et al observations of varying $\alpha _{em}$
predict violations at the $10^{-13}$ level but the Reinhold et al.
observations of varying $\mu $ lead us to expect violations at the $10^{-9}$
level.

In any study of varying constants, the data used to constrain our theories
which allow variations to occur comes from a number of very different
environments and scales: with densities differing by a factor of $10^{30}$
or more, and spanning some 12 billion years. In order to be able to use all
of the information available we need to know how the results of local
laboratory experiments, terrestrial or solar-system bounds from the Oklo
natural reactor, and from isotope ratios in meteorites, are related to data
coming from astronomical observations on extragalactic scales. This is the
`Local vs. Global' problem for varying-constants. It is an important problem
and yet most commentators invariably assume that the local and cosmological
observations are directly comparable \cite{reviews}. This is strong
assumption and is almost invariably made without any proof. A priori, it is
not obvious that this assumption is true; indeed, in many other theories,
not least that for gravity itself, it is not: we do not expect to be able to
measure the expansion of the Universe by observing an expansion of the
Earth. In this paper we describe the first rigorous proof of why, in \emph{%
almost all} varying-constant theories, local experiments will also `see' any
variations in `constants' which occur on cosmological scales.

\section{General Theory}

Before we can solve the `Local vs. Global' problem, we need to introduce the
general way in which a constant is promoted to become a dynamical quantity
consistent with Einstein's conception of gravity. In general, a constant, $%
\mathbb{C}$, is allowed to vary by associating it with some scalar field or
\textquotedblleft dilaton", $\phi $, i.e. $\mathbb{C}\rightarrow \mathbb{C}%
(\phi )$. We usually assume that the scalar field theory associated with $%
\phi $ has a canonical kinetic structure and the variations of this scalar
field contribute to the spacetime curvature like all other forms of
mass-energy. Variations in $\phi $ must also conserve energy and momentum
and so their dynamics are constrained by a non-linear wave equation of the
form 
\begin{equation}
\square \phi =\sum_{j,k}f_{j,\phi }(\phi )L_{j}(\varepsilon _{k},p_{k}),
\label{gen}
\end{equation}%
where $\phi $ is associated with the variation of one or more `constants', $%
\mathbb{C}_{j},$ via a relation $\mathbb{C}_{j}=f_{j}(\phi )$; $f_{j,\phi
}(\phi )=\mathrm{d}f_{j}(\phi )/d\phi $. The $L_{j}(\varepsilon _{k},p_{k})$
are some linear combinations of the density, $\varepsilon _{k}$, and
pressure, $p_{k}$, of the $k^{th}$ species of matter that couples to the
field $\phi $. Included in this formulation are all standard theories for
varying constants, like those for the variation of the Newtonian gravitation
'constant' $G$, $\alpha _{em}$, and the electron-proton mass ratio, as
described in refs. \cite{bd, bek, bsm, bm}.

For local observations to be directly comparable with cosmological ones we
need to know the conditions under which 
\begin{equation*}
\dot{\phi}(\vec{x},t)\approx \phi _{c}(t)
\end{equation*}%
to some specified precision, with $\vec{x}$ taking values in the solar
system, where the subscript $c$ labels the large-scale cosmological value of
the field $\phi $. The validity of this approximate equality, the accuracy
to which it holds, and the accompanying preconditions needed to support its
validity are the subject of the rest of this paper.

Prior to the onset of the matter era the universe is homogeneous to a very
high precision inside the horizon. Any inhomogeneities that do exist, and
the evolution of $\phi $ within them, can be consistently and accurately
described by linear perturbation theory and $\dot{\phi}\approx \dot{\phi}_{c}
$ holds. But the study of the evolution of "constants" becomes
mathematically challenging when linear theory breaks down and the
inhomogeneities become non-linear. This only starts to occur during the
matter era; at these epochs it is an acceptable approximation to consider
the Universe to be comprised of only pressureless dust (baryonic and dark
matter), density $\varepsilon $, and some cosmological constant, $\Lambda $.
We usually expect that the scalar field $\phi $ will couple only to some
fraction of the total dust density; for example, in varying-$\alpha _{em}$
theories it couples to the fraction that feels the electromagnetic force,
and in varying $m_{e}$ theories it couples only to the electron density \cite%
{bm}. We will assume, as is almost always the case, that the fraction of
matter to which it couples is approximately constant during the epoch of
interest. Under these simplifications eqn. \ref{gen} reduces to: 
\begin{equation*}
-\square \phi =B_{,\phi }(\phi )\kappa \varepsilon +V_{,\phi }(\phi )
\end{equation*}%
where $\kappa =8\pi $, $c=G=1$, $V(\phi )$ is the dilaton potential, and $%
B_{,\phi }(\phi )$ is the effective dilaton-to-matter coupling. In what
follows we shall assume that the varying-constant evolves according to the
above conservation equation, which is certainly true of Brans-Dicke theory,
BSBM and BM varying-$\mu $ theory. In what follows we will further assume
that the cosmological value of $\phi $, denoted by $\phi _{c}$, is
sufficiently far away from any extrema of the matter coupling, $B(\phi )$.
We also demand that $V_{,\phi }(\phi )$ not be too large. Our conditions on $%
B(\phi )$ and $V(\phi )$ are summarised as follows: 
\begin{equation*}
\left\vert \frac{B_{,\phi \phi }(\phi _{c})(\phi (\vec{x},t)-\phi _{c}(t))}{%
B_{,\phi }(\phi _{c})}\right\vert \ll 1,\qquad |V_{,\phi }(\phi (\vec{x}%
,t))|\lesssim \Lambda ,
\end{equation*}%
\noindent for all values of $\phi $ within the range of interest (i.e. those
that can be reached from the evolution of some given initial data). The
condition on the matter coupling is usually equivalent to $|B_{,\phi \phi
}(\phi _{c})|\ll 1$ and $B_{,\phi }(\phi _{c})\neq 0$. The condition on the
potential must hold for $\phi =\phi _{c}$ to prevent the varying
\textquotedblleft constant" evolving at an unphysically fast rate
cosmologically; the assumption that holds everywhere will then be valid
provided that $V_{,\phi }(\phi )$ is suitably flat. As a result of this
final assumption our results will \emph{not} apply to Chameleon field
theories, \cite{cham}. With the major exception of theories with
"Chameleonic" behavior, our model includes almost all physically viable
proposals for varying-constant theories. Our results are also applicable to
any scalar-field theory, not just those that describe varying-constants,
provided that the scalar satisfies a conservation equation of the above
form. For a general matter distribution the dilaton conservation equation is
a second-order, non-linear PDE, and there is no reason to suspect that it
should be easily solvable, or indeed analytically solvable at all. Even
numerical calculations will generally be difficult to set-up and control.
Cosmologically, we assume homogeneity and isotropy which leads us to a FRW
background and to a solution for $\phi =\phi _{c}(t)$. Under these
specifications the conservation equation for $\phi $ reduces to an ODE in
time, and can be solved. The other scenario in which it reduces to an ODE is
near a spherically symmetric, static body which couples to the dilaton
strongly enough so that any temporal gradients of $\phi $ are negligible
compared to the spatial ones. In these cases we can easily find the
leading-order static mode of $\phi $, but to find the temporal evolution of $%
\phi $ we need to enforce the boundary condition that it match up to its
cosmological value at large distances. The central technical problem is that
the local, static solution was found under the assumption that the temporal
derivatives were negligible, and at infinity this is no longer the case.
Indeed, to get to a region of space where we know $\phi \approx \phi _{c}(t)$
we must certainly pass through some zone where the temporal and spatial
gradients of $\phi $ are of comparable magnitude. As soon as we reach this
zone, the assumptions under which the local solution was derived break down.
In short: we cannot consistently apply the boundary condition at infinity to
the approximate local solution since spatial infinity is far outside the
range of validity of that approximation. We shall express this idea more
formally below and see that it is associated with the fact that the local
asymptotic approximation is not \emph{uniformly valid}. To circumvent this
problem, created by the presence of multiple length scales, we will use the
method of matched asymptotic expansions.

\section{Matched asymptotic expansions}

Second-order, non-linear PDEs are difficult (and often impossible) to solve
exactly. However, if one can identity some small parameter, $\delta $, in
the problem then it is usually possible to find an expansion in $\delta $
which is formally asymptotic to the solution in the vicinity of some fixed
point. An approximation $\sum^{M}f_{n}(x)\gamma _{n}(\delta )$ is asymptotic
to a function $f(x,\delta )$ as $\delta \rightarrow 0$ iff 
\begin{equation*}
\frac{f(x,\delta )-\sum^{M}f_{n}(x)\gamma _{n}(\delta )}{f_{M}(x)\gamma
_{M}(\delta )}\rightarrow 0\text{ }\mathrm{as}\;\delta \rightarrow 0,
\end{equation*}%
for fixed $x$. If this definition holds for all $M$ then we write: 
\begin{equation*}
f(x,\delta )\sim \sum^{\infty }f_{n}(x)\gamma _{n}(x),
\end{equation*}%
and $\sum^{\infty }f_{n}(x)\gamma _{n}(x)$ is an \emph{asymptotic expansion}
of $f(x,\delta )$ as $\delta \rightarrow 0$ for fixed $x$. The sum here is a
formal sum, since in general it will not converge; however, as a result of
the defining property of asymptotic expansions we will in general need only
the first few terms of the sum to obtain a very good approximation to $%
f(x,\delta )$ at $x$. Asymptotic expansions are unique for each $x$, but it
is also clear that an expansion that is asymptotic to $f(x,\delta )$ for
some range of $x$, with $x\sim \mathcal{O}(1)$ say, will not in general be
valid in some other range of $x$, usually $x\sim \mathcal{O}(1/\delta )$ or $%
x\sim \mathcal{O}(\delta )$. In these cases the expansion is said to be not 
\emph{uniformly valid}. If an expansion has arisen as an approximation to
the solution of a PDE and is not uniformly valid, then the PDE is said to
exhibit \emph{singular} behaviour. Such behaviour is often associated with
the presence of two or more very different length or times scales in the
problem. This is precisely the case in the `Local vs. Global' cosmological
problem, where the length scale of the local inhomogeneity is very much
smaller than the Hubble scale which defines the cosmological background.

We can proceed with such problems by constructing two (or more) asymptotic
approximations to the solutions which are valid for different ranges of $x$,
e.g. for $x\sim \mathcal{O}(1)$ and $x/\delta =\xi \sim \mathcal{O}(1)$,
with 
\begin{eqnarray}
&f(x,\delta )\sim \sum_{n=0}^{Q}f_{n}(x)\delta _{n}\;&\mathrm{as}\;\delta
\rightarrow 0,\;x\;\mathrm{fixed},  \label{innersoln} \\
&f(x,\delta )\sim \sum_{n=0}^{P}g_{n}(\xi )\delta _{n}\;&\mathrm{as}\;\delta
\rightarrow 0,\;\xi =x/\delta \;\mathrm{fixed},  \label{outer}
\end{eqnarray}%
and solving the PDE order by order in $\delta $ for both expansions w.r.t.
to some boundary conditions. We call expansion (\ref{outer}) the \emph{outer
solution}, and (\ref{innersoln}) the \emph{inner solution}. The inner
expansion is not uniformly valid in the region $\xi =\mathcal{O}(1/\delta )$%
, as the outer one is not valid where $x=\mathcal{O}(\delta )$. Because of
these restrictions on the size of $x$, we will only be able to apply a
subset of the boundary conditions to each expansion; in general, we will
therefore be left with unknown coefficients in our asymptotic
approximations. This ambiguity can be lifted if there is some intermediate
region, e.g. $x\sim \mathcal{O}(\delta ^{1/2})$ where they are both valid,
by appealing to the uniqueness of asymptotic expansions and matching the
inner and outer solutions there. In this way we can effectively apply all
boundary conditions to both solutions. This is the method of matched
asymptotic expansions (MAEs). Its application to problems in general
relativity was pioneered by Burke, Thorne and D'Eath \cite{Death} in the
1970s. For a fuller account of MAEs we refer the reader to refs. \cite%
{shawbarrow1} and \cite{hinch}.

\section{Geometrical Set-Up}

The experimental bounds on the permitted level of violations of the WEP due
to the presence of light scalar fields demand that the dilaton field couples
to matter much less strongly than gravity, so 
\begin{equation*}
|B_{,\phi }|\ll 1.
\end{equation*}%
As a result, the dilaton field is only weakly coupled to gravity, and so its
energy density and motion create metric perturbations which have a
negligible effect on the expansion of the background universe. This feature
allows us to consider the dilaton evolution on a fixed background spacetime.
In this work we go further than we did in ref. \cite{shawbarrow1} and
consider not only the extent to which condition $\dot{\phi} \approx \dot{\phi%
}_c$ is satisfied near the surface of some spherical virialised over-density
of matter, e.g. a the Earth, a star, black-hole, galaxy or galaxy cluster,
but also the degree to which it is valid \emph{during} the collapse of an
over-dense region. We will, however, treat the two cases separately.

In the first case, we shall refer to the virialised over-density as our
`star' and take it to have mass $m$ and radius $R_{s}$ at some time of
interest $t=t_{0}$. Although we require that the `star' itself be spherical,
we do not demand that the background spacetime possess any symmetries. We do
require however that, at $t=t_{0}$, the metric is approximately
Schwarzschild, with mass $m$, inside some closed region of spacetime bounded
by a surface at $r=R_{s}$; this region is called the \emph{interior}. The
metric for $r<R_{s}$ is left unspecified. We allow for the possibility that $%
r=R_{s}$ is a black-hole horizon.

In the second case, we only consider the case where the spacetime is
spherically-symmetric, label the mass of the collapsing region by $m$, and
assume that its spatial extent is small compared to the Hubble scale. We
also demand there are no black-hole horizons in the interior of the
collapsing region. By using the results of the first case, however, we can,
in some cases, allow for the formation of a horizon.

In both cases we demand that:

\begin{itemize}
\item Asymptotically, the metric must approach FRW and the whole spacetime
should tend to the FRW metric in the limit $m\rightarrow 0$.

\item The spacetime is approximately FRW in some open region that extends to
spatial infinity, this is called the \emph{exterior}.
\end{itemize}

We are concerned with spacetimes where the matter is a pressureless dust of
density $\varepsilon $, with cosmological constant, $\Lambda $. We further
require that the motion of the dust particles be geodesic. In the
spherically-symmetric case, all such solutions to Einstein's equations with
matter fall into the Tolman-Bondi class of metrics (for a review of these
and other inhomogeneous spherically symmetric metrics see ref \cite{kras}),
however when condition of spherical symmetry is dropped, the general
solution is not known. We can simplify our analysis greatly, however, we
specify four further requirements:

\begin{enumerate}
\item The flow-lines of the background matter are non-rotating. This implies
that the flow-lines are orthogonal to a family of spacelike hypersurfaces, $%
S_{t}$.

\item Each of the surfaces $S_{t}$ is conformally flat.

\item The Ricci tensor for the hypersurfaces $S_{t}$, ${}^{(3)}R_{ab}$, has
two equal eigenvalues.

\item The shear tensor, as defined for the pressureless dust background, has
two equal eigenvalues.
\end{enumerate}

These conditions are automatic if spherical symmetry is required, and in
general they specify the Szekeres-Szafron class of solutions, \cite{szek,
szafron}, of which the Tolman-Bondi solutions, \cite{tolbondi, lem}, are the
spherically symmetric limit. We require the inhomogeneity to be of finite
spatial extent, this limits us to consider only the quasi-spherical Szekeres
solutions, which are described by the metric: 
\begin{equation}
\mathrm{d}s^{2}=\mathrm{d}t^{2}-\frac{\left( 1+\nu _{,R}R\right)
^{2}R_{,r}^{2}\mathrm{d}r^{2}}{1-k(r)}-R^{2}e^{2\nu }\left( \mathrm{d}x^{2}+%
\mathrm{d}y^{2}\right) ,  \label{szekmetric}
\end{equation}%
\noindent where $\nu _{,R}:=\nu _{,r}/R_{,r}$ and

\begin{equation*}
e^{-\nu }=A(r)(x^{2}+y^{2})+2B_{1}(r)x+2B_{2}(r)y+C(r),
\end{equation*}

\begin{equation*}
AC-B_{1}^{2}-B_{2}^{2}=\tfrac{1}{4},
\end{equation*}%
and: 
\begin{equation*}
R_{,t}^{2}=-k(r)+2M(r)/R+\frac{1}{3}\Lambda R^{2}.
\end{equation*}%
In this quasi-spherically symmetric subcase of the Szekeres-Szafron
spacetimes the surfaces of constant curvature, $(t,r)=const$, are 2-spheres 
\cite{szek2}; however, they are not necessarily concentric. These 2-spheres
have surface area $4\pi R^{2}$, and so we deem $R$ to be the \emph{physical
radial coordinate}. In the limit $\nu _{,r}\rightarrow 0$, the $(t,r)=const$
spheres becomes concentric. We can make one further coordinate
transformation so that the metric on the surfaces of constant curvature, $%
(t,r)=const$, is the canonical metric on $S^{2}$ i.e. $\mathrm{d}\theta
^{2}+\sin ^{2}\theta \mathrm{d}\phi ^{2}$: 
\begin{eqnarray}
x\rightarrow X &=&2\left( A(r)x+B_{1}(r)\right) ,  \notag \\
y\rightarrow Y &=&2\left( A(r)y+B_{2}(r)\right) ,  \notag
\end{eqnarray}%
\noindent where $X+iY=e^{i\varphi }\cot \theta /2$. This yields 
\begin{equation*}
-\nu _{,r}|_{x.y}=\frac{\lambda _{z}(X^{2}+Y^{2}-1)+2\lambda _{x}X+2\lambda
_{y}Y}{X^{2}+Y^{2}+1}=\lambda _{z}(r)\cos \theta +\lambda _{x}(r)\sin \theta
\cos \varphi +\lambda _{y}(r)\sin \theta \sin \varphi ,
\end{equation*}%
\noindent where we have defined: 
\begin{equation*}
\lambda _{z}(r):=\frac{A^{\prime }}{A},\qquad \lambda _{x}(r):=\left( \frac{%
2B_{1}}{A}\right) ^{\prime }A,\qquad \lambda _{y}(r):=\left( \frac{2B_{2}}{A}%
\right) ^{\prime }A.
\end{equation*}%
With this choice of coordinates, the local energy density of the dust
separates uniquely into a spherical symmetric part, $\varepsilon _{s}$, and
and a non-spherical part, $\varepsilon _{ns}$: 
\begin{equation*}
\varepsilon =\varepsilon _{s}(t,R)+\varepsilon _{ns}(t,R,\theta ,\varphi ),
\end{equation*}%
\noindent where: 
\begin{eqnarray}
\kappa \varepsilon _{s} &=&\frac{2M_{,R}}{R^{2}}, \\
\kappa \varepsilon _{ns} &=&-\frac{R\nu _{,R}}{1+\nu _{,R}R}\cdot \left( 
\frac{2M}{R^{3}}\right) _{,R}.
\end{eqnarray}%
We define $M_{,R}=M_{,r}/R_{,r}$. We use the remaining freedom to choose $r$
to demand that $r=R$ at $t=t_{0}$.

In the virialised case, we follow the conventions of our earlier papers and
write $M(r):=m+Z(r)$. where $m$ is the gravitational mass of our `star'. In
the spherically-symmetric case $\kappa \varepsilon _{ns}=0$ and the metric
is of Tolman-Bondi form: 
\begin{equation*}
\mathrm{d}s^{2}=\mathrm{d}t^{2}-\frac{R_{,r}^{2}}{1-k(r)}-R^{2}\left\{ 
\mathrm{d}\theta ^{2}+\sin ^{2}\theta \mathrm{d}\phi ^{2}\right\} .
\end{equation*}

\section{Virialised Case}

In ref. \cite{shawbarrow1} we considered whether $\dot{\phi}(\vec{x}%
,t)\approx \dot{\phi}_{c}(t)$ near the surface of some virialised
over-density of matter, which might be a planet, a black-hole, star, or
cluster of galaxies. In general, to specify initial data for the
Szekeres-Szafron solution we must give both the energy density on some
initial hypersurface, $\kappa \varepsilon $, and the spatial curvature of
that hypersurface (given by $k(r)$). In \cite{shawbarrow1} we considered the
two sub-cases of the full Szekeres-Szafron metric, compatible with our
geometric set-up, where the solution is completely specified by giving the
energy density, $\kappa \varepsilon $. :

\begin{itemize}
\item The `Gautreau' case: the hypersurfaces $t=const$ are spatially flat, $%
k(r)=0$. In this case the big-bang is not simultaneous along the past world
lines of all geodesic observers; i.e. it does not occur everywhere for a
single value of $t$, at $t=0$ say. In these cases the flow lines of matter
move out of our `star' - and the mass of `star' decreases of time.

\item The simultaneous big-bang case: the big-bang singularity occurs at $%
t=0 $ for all geodesic observers. In these cases the flow lines of matter
move into the `star' - and its mass increases with time.
\end{itemize}

The first of these cases is simpler to analyse but the second is more
physically reasonable, since we expect the gravity to pull matter onto our
star rather than expel it. For this reason we will only explicitly consider
the simultaneous big-bang case in this paper. The results for the Gautreau
case are very similar and the simultaneity of the big bang is not a
significant factor for the late-time evolutionary problem that we are
considering.

The `interior region', which is immediately outside the surface of the
`star', is approximately Schwarzschild, and so an intrinsic interior length
scale, $L_{I}$, of a sphere centred on the Schwarzschild mass with surface
area $4\pi R_{s}^{2}$ is defined by the Riemann invariant: 
\begin{equation}
L_{I}=\left( \tfrac{1}{12}R_{abcd}R^{abcd}\right) ^{-1/4}=\frac{R_{s}^{3/2}}{%
\left( 2m\right) ^{1/2}}.  \label{invar}
\end{equation}%
In the asymptotically FRW, or exterior, region, the intrinsic length scale
is proportional to the inverse root of the local energy density: $1/\sqrt{%
\kappa \varepsilon _{c}+\Lambda }$, where $\varepsilon _{c}$ is the total
cosmological energy density of matter. We shall assume that the FRW region
is approximately flat ($k=0$), and we define a length scale appropriate for
this exterior region at epoch at $t=t_{0}$ equal to the Hubble radius, which
is defined by the inverse Hubble parameter at that time: 
\begin{equation*}
L_{E}=1/H_{0}.
\end{equation*}%
For realistic models $L_{E}\gg L_{I}$ and so we define $\delta $ to be a
small parameter given by

\begin{equation*}
\delta =L_{I}/L_{E}.
\end{equation*}

We assume that the whole spacetime metric is Szekeres-Szafron and define
dimensionless coordinates appropriate to both the interior and exterior near
some epoch of interest at time $t=t_{0}$. In the interior: 
\begin{equation*}
T=L_{I}^{-1}(t-t_{0}),\quad \xi =R_{s}^{-1}R,
\end{equation*}%
\noindent where $R$ is the physical radial coordinate; $T$ and $\xi $ are $%
\mathcal{O}(1)$ in the interior and we take the ratio $2m/R_{s}$ to be
fixed. It is also helpful to define 
\begin{equation*}
\eta =\left( \xi ^{3/2}-3T/2\right) ^{2/3};\text{ }R_{s}\eta =r+\mathcal{O}%
(\delta ^{q},\delta ^{2/3}).
\end{equation*}%
In the exterior we define: 
\begin{equation*}
\tau =H_{0}t,\quad \rho =H_{0}r,
\end{equation*}%
where $r$ is the unphysical radial labelling coordinate used in the metric %
\ref{szekmetric}. We define the \emph{interior limit} by $\delta \rightarrow
0$ for fixed $T$ and $\xi $, and the \emph{exterior limit} by $\delta
\rightarrow 0$ with $\tau $ and $\rho $ held fixed.

\subsection{The Exterior Limit}

According to our prescription that the metric be FRW to zeroth order in $%
\delta$, we write

\begin{equation*}
H_{0}Z(\rho )\sim \frac{1}{2}\Omega _{m}\rho ^{3}+\delta
^{p}z_{1}(r)+o(\delta ^{p}),
\end{equation*}
and

\begin{equation*}
H_{0}^{-1}\lambda _{i}\sim \delta ^{s}l_{i}(\rho )+o(\delta ^{s}),
\end{equation*}%
where $s$ and $p$ are positive numbers which depend on the particular form
of the initial matter distribution. The exterior expansion of $k(r)$ can be
found using the exact solutions\ for the Szekeres metrics with cosmological
constant \cite{JBJSS}.

Since $H_{0}^{-2}\left( \frac{2M}{R^{3}}\right) _{,R}\sim \mathcal{O}(\delta
^{p},\delta )$, we have that: $H_{0}^{-2}\kappa \varepsilon _{ns}\sim 
\mathcal{O}(\delta ^{p+s},\delta ^{1+s})$ whereas $H_{0}^{-2}\kappa
\varepsilon _{s}\sim \mathcal{O}(\delta ^{p},\delta )$. Thus, the
non-spherical perturbation to the energy density is always of sub-leading
order compared to the first order in spherical perturbation. The
first-order, non-spherical, metric perturbation appears at $\mathcal{O}%
(\delta ^{s})$; however, this is equivalent to a coordinate transform on $%
(r,\theta ,\varphi )$ and does not source a non-spherically symmetric
physical perturbation to the dilaton evolution at this order. For the
dilaton field, $\phi $, then, the first non-spherical perturbation is always
sourced at subleading order compared to the first spherically symmetric one.

In the exterior we can apply the boundary condition that $\phi \rightarrow
\phi _{c}(t)$ as $r\rightarrow \infty $. In addition to this we also have
the stronger condition that, as $\delta \rightarrow 0,$ the inhomogeneity
should disappear and $\phi \rightarrow \phi _{c}(t)$. Thus to zeroth order
in the exterior $\phi \sim \phi _{c}(t)+\mathcal{O}(\delta ^{p},\delta )$.
Since we are only really interested in the behaviour of $\phi $ in the
interior we do not need to calculate the higher-order terms in the exterior
limit explicitly, we only need to know enough about their behaviour to be
able to perform the matching in some intermediate scaling region. We will
consider that behaviour later.

\subsection{Interior Limit}

To lowest order in the interior region, we write $Z\sim \delta ^{q}R_{s}\mu
_{1}$, and $\lambda _{i}:=\delta ^{q^{\prime }}R_{s}^{-1}b_{i}$, where $%
i=\{x,y,z\}$; $q$ and $q^{\prime }$ are determined by specify a particular
matter distribution. The condition that $\kappa \varepsilon >0$ everywhere
requires $q^{\prime }\geq q$. From the exact solutions we find: 
\begin{equation*}
k(r)\sim \delta ^{2/3}k_{0}\left( 1+\delta ^{q}\mu _{1}(\eta )+o\left(
\delta ^{q}\right) \right) +\mathcal{O}\left( \delta ^{5/3}\right) ,
\end{equation*}%
where $k_{0}(\delta T)=(2m/R_{s})\left( \pi /(H_{0}t_{0}+\delta T)\right)
^{2/3}$. In refs.\cite{shawbarrow1} we sought to remove the effect of the $%
\mathcal{O}(\delta ^{2/3})$ in $k(r)$ by a transformation of the time
coordinate. However, it is not clear that this new time coordinate is
well-defined near a black-hole horizon; we now believe this procedure to
have been technically incorrect (although it did not effect the results). We
correct it in this work by implementing the $\mathcal{O}(\delta ^{2/3})$
correction differently. If $q^{\prime }>q$ then to the next-to-leading
order, we need only consider the spherically-symmetric modes to find
interior expansion of $\phi $. We could also include a non-spherical vacuum
component for $\phi $ at next-to-leading order; however, this will be
entirely determined by a boundary condition on $R=R_{s}$ and the need that
it should vanish for large $R$. To find the leading-order behaviour of the $%
\phi _{,T}$ we need to know $\phi $ at next-to-leading order. Hence, the
only case where we must explicitly consider non-spherically symmetric
effects is when $q^{\prime }=q$, i.e. $\kappa \varepsilon _{ns}=\mathcal{O}%
(\kappa \varepsilon _{s})$. In what follows it is natural to consider the
spherically symmetric and non-spherically symmetric modes of $\phi $
separately.

Before we can solve the $\phi $ equations in the interior limit, we need a
boundary condition at $R=R_{s}$. At leading order we take this to be: 
\begin{eqnarray*}
R_{s}^{2}\left( 1-\frac{2m}{R_{s}}\right) \left. \partial _{R}\phi
_{0}\right\vert _{R=R_{s}}=2mF\left( \bar{\phi}_{0}\right)  && \\
=\int_{0}^{R_{s}}\mathrm{d}R^{\prime }R^{\prime }{}^{2}B_{,\phi }(\phi
_{0}(R^{\prime },t))\kappa \varepsilon (R^{\prime }), &&
\end{eqnarray*}%
The no-hair theorem for black holes implies that $F\left( \bar{\phi}%
_{0}\right) =0$, however for bodies where $2m/R_{s}\ll 1$ we expect $F\left( 
\bar{\phi}_{0}\right) \approx B_{,\phi }(\phi _{c})$. At higher orders we
find the flux $F$ by perturbing the above expression as explained in \cite%
{shawbarrow1}. The zeroth-order mode of $\phi $ in the interior is then
found to be: 
\begin{equation*}
\phi ^{(0)}=\phi ^{(0)}(T,\xi ):=\phi _{e}\left( \delta T\right) +F\left( 
\bar{\phi}_{0}\right) \ln \left( 1-\frac{2m}{R_{s}\xi }\right) .
\end{equation*}%
The matching procedure gives $\phi _{e}\left( \delta T\right) =\phi _{c}(t)$.

\subsubsection{Spherically Symmetric Perturbations}

In \cite{shawbarrow1} we considered the spherically symmetric perturbations
of $\phi $ that occur at order $\delta ^{q}$ and at order $\delta $. Here,
we will also consider the perturbations at order $\delta ^{2/3}$. The
perturbation at order $\delta $ is sourced by the $\square _{0}\phi
_{e}(\delta T)$ term, where $\square _{0}$ is the d'Alembertian of the
Schwarzschild metric. This effect of perturbation acts like a drag term, and
is equivalent to the change $\phi _{e}(\delta T)\rightarrow \phi _{e}(\delta
T^{\prime })$ with: 
\begin{equation*}
T^{\prime }=T-2\left( \frac{2m}{R_{s}}\right) ^{3/2}\left( \sqrt{\frac{%
R_{s}\xi }{2m}}-\ln \left\vert 1+\sqrt{\frac{2m}{R_{s}\xi }}\right\vert
\right) +C,
\end{equation*}%
where $C$ is some constant which is determined by the matching process. When
the matching is performed one finds: 
\begin{equation*}
T^{\prime }\sim T+\left( \frac{2m}{R_{s}}\right) \int_{\infty }^{\xi }%
\mathrm{d}\xi ^{\prime }\frac{\xi _{,T}(\xi ^{\prime },T)\xi ^{\prime 2}-\xi
_{,T}(1,T)}{\xi ^{\prime 2}(1-k(\xi ^{\prime },t)-(2m/R_{s})^{1/2}\xi
_{,T}^{2}(\xi ^{\prime },T))}
\end{equation*}%
where $\xi _{,T}(\xi ,T)=R_{,t}(R=R_{s}\xi ,t=t_{0}+L_{I}T)$. When the
`star' is actually a black hole, the above expression reproduces, to leading
order in $\delta $, the result found by Jacobson \cite{jacobson}. As for the 
$\mathcal{O}(\delta ^{2/3})$ correction to the metric coming from $k(r)$, we
find that it only effects $\phi $ at order $\mathcal{O}(\delta ^{5/3},\delta
^{q+2/3})$; as such, they can be ignored since they are always smaller than
the effects that we have included.

In \cite{shawbarrow1} we calculated the order $\delta ^{q}$ correction to $%
\phi $, $\phi ^{(q)}(T,\xi )$. In cases where the surface of our `star' is
far outside its Schwarzschild horizon ($2m/R_{s}$), $\phi _{I}^{(q)}$ is
given by: 
\begin{equation*}
\phi _{I}^{(q)}\sim \frac{2m}{R_{s}}B_{,\phi }\left( \phi _{c}\right) \left(
\int^{\eta }\mathrm{d}\eta ^{\prime }\frac{\mu _{1}\left( \eta ^{\prime
}\right) _{,\eta }}{\xi (\eta ^{\prime },T)}-\frac{\mu _{1}\left( \eta
\right) }{\xi }+\left( 1-\frac{F(\bar{\phi}_{0})}{B(\phi _{c})}\right) \frac{%
\mu _{1}\left( \eta (\xi =1,T\right) }{\xi }+D(T)\right) +\mathcal{O}\left(
\left( \frac{2m}{R_{s}}\right) ^{2}\right) .
\end{equation*}%
In the cases where $2m/R_{s}\approx 1$ it was not possible to find a closed
analytical expression for $\phi _{I}^{(q)}$; however, from the its equation
of motion, it can be easily seen that $\phi _{I}^{(q)}$ will be of the same
order of magnitude as the above expression. The function $D(T)$ is a
constant of integration, and it must be found via the matching procedure.
Before we perform this matching, we will consider the non-spherically
symmetric modes.

\subsubsection{Non-Spherically Symmetric Perturbations}

Non-spherically symmetric modes in the interior approximation to $\phi $
will be sourced at order $\delta ^{q^{\prime }}$, where $q^{\prime }\geq q$.
We studied these modes and found that, to order $\delta ^{q}$, they only
have a dipole moment. We write the order $\delta ^{q^{\prime }}$
non-spherically symmetric, modes as $\delta ^{q^{\prime }}\phi
_{I}^{(q^{\prime })}$ where: 
\begin{equation*}
\phi _{I}^{(q^{\prime })}:=\phi _{I}^{(q^{\prime })z}(\xi ,T)\cos \theta
+\phi _{I}^{(q^{\prime })x}(\xi ,T)\sin \theta \cos \varphi +\phi
_{I}^{(q^{\prime })y}(\xi ,T)\sin \theta \sin \varphi .
\end{equation*}%
As before, when $2m/R_{s}\ll 1,$ we can find analytic expressions for these
modes: 
\begin{eqnarray}
\phi _{I}^{(q^{\prime })i} &\sim &-\frac{2m}{R_{s}}B_{,\phi }\left( \phi
_{I}^{0}\right) \xi \int^{\eta }\mathrm{d}\eta ^{\prime }\frac{b_{i}(\eta
^{\prime })}{\xi ^{^{\prime }2}}+\frac{2m}{R_{s}}B_{,\phi }\left( \phi
_{I}^{0}\right) \frac{1}{\xi ^{2}}\int_{\xi =1}^{\eta }\mathrm{d}\eta
^{\prime }\xi ^{\prime }b_{i}(\eta ^{\prime }) \\
&-&\frac{2m}{R_{s}}F\left( \bar{\phi}_{0}\right) \frac{1}{\xi ^{2}}\int_{\xi
=1}^{\eta }\mathrm{d}\eta ^{\prime }b_{i}(\eta ^{\prime })\xi ^{\prime }+%
\frac{C_{i}(T)}{\xi ^{2}}+D_{i}(T)\xi +\mathcal{O}((2m/R_{s})^{2}),  \notag
\end{eqnarray}%
where $i=x,\,y\,z$. When $2m/R_{s}\approx 1$ the above expression can be
seen as an order of magnitude estimate for the $\phi _{I}^{(q^{\prime })i}$.
The $D_{i}(T)$ and $C_{i}(T)$ are constants of integration, with $D_{i}(T)$
determined by the matching procedure. The value of $C_{i}(T)$ should be set
by a boundary condition on $R=R_{s}$. We cannot specify $C_{i}(T)$ exactly
without further information about the interior of our `star' in $R<R_{s}$.
If we assume that the prescription for the sub-leading order boundary
condition given above is correct then we find: 
\begin{eqnarray}
\partial _{\xi }\phi _{I}^{(q^{\prime })i}|_{\xi =1} &\sim &-\frac{2m}{R_{s}}%
\left. \frac{b_{i}}{\eta ^{1/2}}\right\vert _{\xi =1}F\left( \bar{\phi}%
_{0}\right) +\mathcal{O}((2m/R_{s})^{2})  \notag \\
\Rightarrow C_{i} &=&-\frac{m}{R_{s}}B_{,\phi }\left( \phi _{I}^{0}\right)
\int^{\xi =1}\mathrm{d}\eta ^{\prime }\frac{b_{i}(\eta ^{\prime })}{\xi
^{^{\prime }2}}+\tfrac{1}{2}D_{i}  \notag
\end{eqnarray}%
From now on, we set $C_{i}=0,$ for simplicity. Even when this is not exactly
satisfied, we do not expect the magnitude of $C_{i}$ or $C_{i,T}$ to be
larger than any of the other terms in $\phi _{I}^{(1)i}$ or $\phi
_{I,T}^{(1)i}$, respectively.

\subsection{Validity of Matching Procedure}

Before we can apply the matching procedure, we must ensure that it is
applicable to our problem i.e. that there exists some intermediate region
where both the interior and exterior approximations are simultaneously
valid. We considered these conditions in ref.\cite{shawbarrow1}. We define
coefficients $n>0$ and $d_{i}>0$ by $\mu _{1}\left( \eta \right) \sim \eta
^{n}$ and $b_{i}\left( \eta \right) \propto \eta ^{d_{i}}$ as $\eta
\rightarrow \infty $ respectively. Writing $H_{0}^{-1}\lambda _{i}\sim
\delta ^{p_{i}^{\prime }}l_{i}(\rho )$, we also define coefficients $m$ and $%
f_{i}$ by $z_{1}(\rho )\sim \rho ^{m}$ and $l_{i}(\rho )\propto \rho
^{-f_{i}}$ as $\rho \rightarrow 0$. For both the exterior and interior to be
simultaneously valid in some intermediate scaling region where $\eta ,\chi
,\xi \sim \delta ^{-\alpha }$ with $0<\alpha <1$, we need there to exist
some $\alpha $ such that: 
\begin{eqnarray}
&&\max \left( 0,1-\frac{p}{1-m}\right) <\alpha <\frac{q}{n}  \notag \\
&&\max_{i}(p_{i}^{\prime }+(1-\alpha )(f_{i}+m))>-p\;\mathrm{if}\;p\leq 1, 
\notag \\
&&\max_{i}(p_{i}^{\prime }+(1-\alpha )f_{i})>-1\;\mathrm{if}\;p\geq 1. 
\notag \\
&&\alpha -\max_{i}{q/d_{i}}>0  \notag
\end{eqnarray}%
These conditions can, in almost all cases, be rephrased as: 
\begin{eqnarray}
\lim_{\delta \rightarrow 0}{R^{2}\kappa \Delta \varepsilon } &=&o(1) \\
\lim_{\delta \rightarrow 0}{2(m+Z)/R} &=&o(1)
\end{eqnarray}%
as $\delta \rightarrow 0$, with $L_{I}^{\alpha }L_{E}^{1-\alpha
}(t-t_{0}),L_{I}^{\alpha }L_{E}^{1-\alpha }R$ \ held fixed, for all $\alpha
\in (0,1)$.

\subsection{Matching and Results}

We are interested in time derivatives of the $\phi $ field. From the
expression for $\phi $ in the interior, we see that $t^{\prime
}=L_{I}T^{\prime }$ seems to play the role of a natural time coordinate. At
radii where $2m/R\ll 1$, the interior metric is close to diagonal when
written in $(t^{\prime },R)$ coordinates; it is in this sense a \emph{%
natural time coordinate} for an observer at fixed $R$. In this region, $%
t^{\prime }$ coincides with the standard Schwarzschild time coordinate. As $%
R\rightarrow \infty $, $t^{\prime }\rightarrow t$. We therefore consider $%
\phi _{,t^{\prime }}$. Whenever the required conditions of the previous
section hold, the matching procedure is valid, and we find 
\begin{equation*}
\phi _{,t^{\prime }}(r,t)\approx \phi _{I,t^{\prime }}^{(0)}+\delta ^{q}\phi
_{I,t^{\prime }}^{(q)}+\delta ^{q^{\prime }}\phi _{I,t^{\prime
}}^{(q^{\prime })}+o(\delta ^{q},\delta ^{q^{\prime }})
\end{equation*}%
in the interior, where the $\approx $ sign means that whilst this is not a
formal asymptotic series (since there may be excluded terms that are bigger
than some of the included ones) this is a good numerical estimate since 
\emph{at least one of the included terms will be bigger than all the
excluded ones}. The terms in this expression are given by: 
\begin{eqnarray}
\phi _{I,t^{\prime }}^{(0)} &\sim &\dot{\phi}_{c}(t)+\Delta t(r,t)\ddot{\phi}%
_{c}(t), \\
\Delta t(r,t) &=&\int_{\infty }^{r}\mathrm{d}r^{\prime }R_{,r}\frac{\Delta
(R_{,t})R^{2}-R_{s}^{2}\Delta (R_{,t})|_{R=R_{s}}}{R^{2}(1-k(r^{\prime
})-R_{,t}^{2})}  \label{deltat}
\end{eqnarray}%
where $\Delta R_{,t}=R_{,t}-HR$, 
\begin{equation}
\delta ^{q}\phi _{I,t^{\prime }}^{(q)}\sim -B_{,\phi }(\phi _{c})\left(
\int_{\infty }^{r}\mathrm{d}r^{\prime }\Delta R_{,r}(R_{,t}\kappa
\varepsilon )(r^{\prime },t)+\left( 1-\frac{F(\bar{\phi}_{0})}{B_{,\phi
}(\phi _{c})}\right) \frac{\left( R_{s}^{2}R_{,t}\kappa \Delta \varepsilon
\right) _{R=R_{s}}}{R}\right) +\mathcal{O}\left( \left( \frac{2m}{R_{s}}%
\right) ^{2}\right)   \label{phist}
\end{equation}%
and 
\begin{eqnarray}
\delta ^{q^{\prime }}\phi _{I,t^{\prime }}^{(q^{\prime })} &\sim &-\frac{2}{3%
}B_{,\phi }\left( \phi _{I}^{0}\right) R\int_{\infty }^{r}\mathrm{d}%
r^{\prime }R_{,r}R_{,t}\frac{\kappa \varepsilon _{ns}(r^{\prime },t)}{R}-%
\frac{1}{3}B_{,\phi }\left( \phi _{I}^{0}\right) \frac{1}{R^{2}}%
\int_{R=R_{s}}^{r}\mathrm{d}r^{\prime }R_{,r}R_{,t}R^{2}\kappa \varepsilon
_{ns}(r^{\prime },t)  \label{phinst} \\
&&+\frac{1}{3}F\left( \bar{\phi}_{0}\right) \frac{1}{R^{2}}\int_{R=R_{s}}^{r}%
\mathrm{d}r^{\prime }R_{,r}R_{,t}R^{3}\kappa \varepsilon _{ns}(r^{\prime
},t)-\frac{1}{3}F\left( \bar{\phi}_{0}\right) RR_{,t}\kappa \varepsilon
_{ns}(r,t)+\mathcal{O}\left( \left( \frac{2m}{R_{s}}\right) ^{2}\right) . 
\notag
\end{eqnarray}%
Equations (\ref{phist}) and (\ref{phinst}) are derived for the case $%
2m/R_{s}\ll 1$. For $2m/R_{s}\approx 1$, they are accurate when $R\gg 2m$
and otherwise provide order-of-magnitude estimates for $\delta ^{q}\phi
_{I,t^{\prime }}^{(q)}$ and $\delta ^{q^{\prime }}\phi _{I,t^{\prime
}}^{(q^{\prime })}$ respectively.

We can now evaluate these terms to find out when $\dot{\phi}(\vec{x}%
,t)\approx \dot{\phi}_{c}(t)$ and also state the precision to which this
approximate equality holds. In many cases, however, a lot of the terms in
the above expression are negligible or cancel, and so we can find a more
succinct necessary and sufficient condition for $\dot{\phi}(\vec{x}%
,t)\approx \dot{\phi}_{c}(t)$ to hold. When $2m/R_{s}\gg 1$ we expect $%
F\left( \bar{\phi}_{0}\right) \approx B_{,\phi }\left( \phi _{I}^{0}\right)
\left( 1+\mathcal{O}(2m/R_{s})\right) $, and so: 
\begin{eqnarray}
\phi _{I,t^{\prime }}-\phi _{c,t} &\approx &-B_{,\phi }\left( \phi
_{c}\right) \int_{\infty }^{r}\mathrm{d}r^{\prime }R_{,r}R_{,t}\kappa \Delta
\varepsilon _{s}(r^{\prime },t)-\frac{2}{3}B_{,\phi }\left( \phi _{c}\right)
R\int_{\infty }^{r}\mathrm{d}r^{\prime }R_{,r}R_{,t}\frac{\kappa \varepsilon
_{ns}(r^{\prime },t)}{R}  \label{deltaphiv} \\
&-&\frac{1}{3}B_{,\phi }\left( \phi _{c}\right) RR_{,t}\kappa \varepsilon
_{ns}(r,t)+\Delta t(r,t)\ddot{\phi}_{c}(t).
\end{eqnarray}%
We will refer to this last term as the \emph{drag term}, and it is
responsible for the local value of $\dot{\phi}$ lagging slightly behind the
cosmological one - this effect was first observed by Jacobson in the study
of gravitational memory, \cite{jacobson}. Whenever the cosmological $\phi $
is \emph{not} potential dominated, and our `star' resides in a local
overdensity of matter, the drag term will be negligible compared to the
other terms in this expression. If the potential term dominates the
cosmological evolution then it is possible for the drag term to give the
dominant effect, even if we have an local over-density of matter. But
whenever this happens we always have $|\Delta t\,\ddot{\phi}_{c}/\dot{\phi}%
_{c}|\ll 1$ and so we have $\dot{\phi}(\vec{x},t)\approx \dot{\phi}_{c}(t)$.

Independent of the nature of overdensity, we saw in our previous papers that
potential domination of the cosmological $\phi $ evolution acts only to
strengthen the degree to which $\dot{\phi}(\vec{x},t)\approx \dot{\phi}%
_{c}(t)$.

Even if we ignore the drag term, the above expression for $\phi
_{I,t^{\prime }}-\phi _{c,t}$ is still rather unwieldy. In almost all cases,
integrating over the non-spherically symmetric modes of $\kappa \varepsilon $
in the same way as we did for the spherically symmetric ones only acts to
increase $|\phi _{I,t^{\prime }}-\phi _{c,t}|$. Therefore, we define the
quantity $\mathcal{I}$ by:

\begin{equation}
\mathcal{I}:=B_{,\phi }(\phi _{c})\int_{\infty }^{R}\mathrm{d}R^{\prime }%
\frac{\max_{\theta ,\phi }(\sin \theta \Delta (v\varepsilon ))}{\dot{\phi
_{c}}}\ll 1  \label{suff}
\end{equation}%
where $v$ is the radial velocity of the matter particles (i.e. $v=R_{,t}$).
When the background spacetime is Szekeres-Szafron we have that: 
\begin{equation*}
\left\vert \frac{\phi _{,t^{\prime }}-\phi _{c,t}}{\phi _{c,t}}\right\vert
\lesssim \mathcal{I}
\end{equation*}%
with equality in the spherically symmetric case. The strong inequality $%
\mathcal{I}\ll 1$ is therefore a sufficient condition for $\dot{\phi}(\vec{x}%
,t)\approx \dot{\phi}_{c}(t)$, and the value of $\mathcal{I}$ gives a
measure of the amount by which $\phi _{,t^{\prime }}$ and $\phi _{c,t}$
differ. In addition to having shown this for cases where the background is
Szekeres-Szafron, and that matching conditions hold, we also conjecture
that, even if the matching conditions formally fail, and for other classes
of spacetime, that $\mathcal{I}\ll 1$ is a sufficient condition for $\dot{%
\phi}(\vec{x},t)\approx \dot{\phi}_{c}(t)$.

\section{Collapsing Case}

When a spacetime undergoes gravitational collapse it is possible for a
black-hole to form inside the collapsing region. In the Tolman-Bondi model a
black-hole horizon appears when $2M(r)/R=1$. If we have $\kappa \Delta
\varepsilon R^{2}=(\kappa \varepsilon -\kappa \varepsilon _{c})R^{2}\ll 1$
outside the horizon then we can apply the results of the previous section,
taking the surface of our `star' to be the black-hole horizon. This is also
true for any virialised region in the interior of the collapsing region, not
just for black holes.

The results of the previous section can also be extended to the case where
the collapsing interior region has \emph{no} central black-hole or
virialised region. For simplicity we consider only spherically-symmetric,
dust-plus-$\Lambda $ cosmologies i.e. Tolman-Bondi models. We do \emph{not}
require the big bang to be simultaneous for all observers. This extension
requires that curvature of the interior spacetime be in some sense weak so
that the metric is Minkowski to zeroth order. We require: 
\begin{equation*}
R^{2}\kappa \Delta \varepsilon \ll 1,\qquad 2\Delta M/R\ll 1
\end{equation*}%
everywhere; $\Delta \varepsilon =\epsilon -\epsilon _{c}$ and $\Delta M=M-%
\frac{1}{6}\kappa \epsilon _{c}$. in the interior. In the exterior we
assume, as before, that the spacetime is FRW to zeroth order. It is clear
that in this model the following parameters, $\delta _{1}$ and $\delta _{2}$
are \emph{everywhere} small: 
\begin{eqnarray}
\delta _{1}(R,t) &=&R^{2}(\kappa \varepsilon -\kappa \varepsilon _{c})=\frac{%
2M_{,r}}{R_{,r}}-3\Omega _{m}H^{2}R^{2},  \notag \\
\delta _{2}(R,t) &=&\frac{2M}{R}-\Omega _{m}H^{2}R^{2}.  \notag
\end{eqnarray}%
In addition, the following parameter, $\delta _{3},$ is small in interior
region but $O(1)$ in the exterior: 
\begin{equation*}
\delta _{3}(R,t)=H^{2}R^{2}.
\end{equation*}%
In the interior $\delta _{3}\ll \delta _{1},\delta _{2}$. For the purposes
of our asymptotic expansions we treat and $\delta _{1}$ and $\delta _{2}$ as
being of the same order. For the interior to be collapsing we need $k(r)>0$.
The condition that $R_{,t}^{2}=-k(r)+\delta _{2}+(\Omega _{m}+\Omega
_{\lambda })\delta _{3}>0$ implies that $k(r)\lesssim \mathcal{O}(\delta
_{2})$, and $R_{,t}^{2}\sim \mathcal{O}(\delta _{2})$ in the interior.

We will perform the matching in an intermediate region where $\delta
_{1}\sim \delta _{2}\sim \delta _{3}\ll 1$. It is clear that with these
definitions that such an intermediate region must always exist.

\subsection{The Interior}

In the interior we write the metric as: 
\begin{equation*}
\mathrm{d}s^{2}=\frac{(1-\delta _{2}+\delta _{3}(\Omega _{m}+\Omega
_{\lambda }))\mathrm{d}t^{2}}{1-k(r)}+\frac{2R_{,t}\mathrm{d}R\mathrm{d}t}{%
1-k(r)}-\frac{\mathrm{d}R^{2}}{(1-k(r))}-R^{2}\{\mathrm{d}\theta ^{2}+\sin
^{2}\theta \mathrm{d}\varphi ^{2}\},
\end{equation*}%
which is flat spacetime to lowest order in the $\delta _{i}$. The dilaton, $%
\phi $, obeys: 
\begin{equation*}
-R^{2}\square \phi =B_{,\phi }(\phi )(\delta _{1}+3\Omega _{m}\delta
_{3})+V_{,\phi }(\phi )R^{2},
\end{equation*}%
and, in line with our previous assumptions, we have $V_{,\phi }(\phi
)R^{2}\sim \mathcal{O}(\Omega _{\Lambda }\delta _{3})$. We note that $%
\partial _{t}\delta _{1}\sim \mathcal{O}(\delta _{1}^{1/2}\delta
_{2})=o(\delta _{1}^{3/2})$ and so $\delta _{1}$ is quasi-static; as such we
expect $\phi $ to also be quasi-static in the interior. We can solve the
equations for $\phi $ order-by-order in the interior, requiring (as a
boundary condition) that $\phi $ is regular at $R=0$: 
\begin{eqnarray}
\phi \approx \phi _{e}(t) &+&B_{,\phi }(\phi _{e})\int_{C}^{R}\frac{\mathrm{d%
}R^{\prime }}{R^{\prime }}\delta _{2}(R^{\prime },t)+\dot{\phi}%
_{e}(t)\int_{D}^{R}\mathrm{d}R^{\prime }\left( R_{,t}-HR^{\prime }\right) 
\label{intcol} \\
&+&\frac{1}{6}\left[ R^{2}V_{,\phi }(\phi _{e})+3B_{,\phi }\Omega _{m}\delta
_{3}+(\ddot{\phi}_{e}(t)+3H\dot{\phi}_{e}(t))R^{2}\right] +\mathcal{O}\left(
\delta _{1}^{2},\delta _{2}^{2},\delta _{3}^{2},(R\dot{\phi}_{e})^{3},\delta
_{3}(R\dot{\phi}_{e})\right) ,  \notag
\end{eqnarray}%
where $\phi _{e}(t)$ is $\mathcal{O}(1)$ but quasi-static i.e. $R\dot{\phi}%
_{e}=o(\delta _{1}^{1/2},\delta _{2}^{1/2})$; the third term is $\mathcal{O}%
(\delta _{2}^{1/2}R\dot{\phi}_{e})$. Since the above expression is not a
formal asymptotic expansion as such we cannot be sure that the neglected
terms are smaller than \emph{all} of the included terms; indeed we shall see
that the matching ensures the vanishing of the term in $\left[ ..\right] $.
This is because we do not know precisely how the sizes of $\delta _{3}$ and $%
R\dot{\phi}_{e}$ relate to those of $\delta _{1}$ and $\delta _{2}$. What we
do know is that, in the interior, the excluded terms \emph{are} smaller than
at least one of the included terms. The limits $C$ and $D$ as well as $\phi
_{e}(t)$ must be found matching the interior expansion to the exterior one.

\subsection{The Exterior}

In the exterior we define a coordinate $\varrho =R/a(t)\sim r$ where $a(t)$
is the FRW scalar factor. In $(t,\varrho )$ coordinates the metric reads: 
\begin{equation*}
\mathrm{d}s^{2}=\mathrm{d}t^{2}(1+\mathcal{O}(\delta _{2}^{2},(\Delta
k)^{2})+\frac{2\left( R_{,t}-HR\right) a\mathrm{d}\varrho \mathrm{d}t}{1-k(r)%
}-\frac{a^{2}\mathrm{d}\varrho ^{2}}{1-k(r)}-a^{2}\varrho ^{2}\{\mathrm{d}%
\theta ^{2}+\sin ^{2}\theta \mathrm{d}\phi ^{2}\}
\end{equation*}%
where $\Delta k=k(r)-k_{0}\varrho ^{2}$, and $k_{0}=\lim_{r\rightarrow
\infty }k(r)/r^{2}$, $(1-\Omega _{m}-\Omega _{\Lambda
})H^{2}=-k_{0}^{2}/a^{2}$ and $2\left( R_{,t}-HR\right) \sim (-\Delta {k}%
+\delta _{2})/HR$. As $R,\varrho \rightarrow \infty $ and the inhomogeneity
is removed (i.e. $\Delta k,\delta _{1},\delta _{2}\rightarrow 0$) we require
that $\phi \rightarrow \phi _{c}(t)$. As with the virialised case our only
interest in the subleading order behaviour of $\phi $ in the exterior is so
as to match it to the interior approximation. It is only necessary therefore
to consider how the exterior approximation to $\dot{\phi}$ behaves in the
intermediate region where all the $\delta _{i}$ are small. In the
intermediate region the exterior approximation is: 
\begin{equation*}
\phi \sim \phi _{c}(t)+\phi _{s}(\varrho ,t)+B_{,\phi }(\phi
_{c})\int_{\infty }^{R}\frac{\mathrm{d}R^{\prime }}{R^{\prime }}\delta
_{2}(R^{\prime },t)+\dot{\phi}_{c}(t)\int_{\infty }^{R}\mathrm{d}R^{\prime
}(R_{,t}-HR^{\prime })+\mathcal{O}(\delta _{1}^{2},\delta _{2}^{2},\delta
_{3}^{3}),
\end{equation*}%
where $\phi _{s}(\varrho ,t)=o(1)$ is some vacuum mode i.e. $\square
_{FRW}\phi _{s}=0$.

\subsection{Matching and Results}

Now we match the interior and exterior expansions and find that $C=D=\infty $%
, $\phi _{e}(t)=\phi _{c}(t)$ and $\phi _{s}=0$. The requirement that $\phi
_{e}(t)=\phi _{c}(t)$, combined with the cosmological evolution equation for 
$\phi _{c}(t)$ ensure that the term in $\left[ ..\right] $ in eqn. (\ref%
{intcol}) vanishes. We have found that the matching interior approximation
is therefore given by: 
\begin{eqnarray*}
\phi  &\sim &\phi _{c}(t)+B_{,\phi }(\phi _{c})\int_{\infty }^{R}\frac{%
\mathrm{d}R^{\prime }}{R^{\prime }}\delta _{2}(R^{\prime },t)+\dot{\phi}%
_{c}(t)\int_{\infty }^{R}\mathrm{d}R^{\prime }(R_{,t}-HR^{\prime })+\mathcal{%
O}(\delta _{1}^{2},\delta _{2}^{2},\delta _{3}^{3}) \\
&\sim &\phi _{c}(t^{\prime }=t+\Delta t)+B_{,\phi }(\phi _{c})\int_{\infty
}^{R}\frac{\mathrm{d}R^{\prime }}{R^{\prime }}\delta _{2}(R^{\prime },t)+%
\mathcal{O}(\delta _{1}^{2},\delta _{2}^{2},\delta _{3}^{3}).
\end{eqnarray*}%
where the lag in the time coordinate, $\Delta t$, is given by: 
\begin{equation*}
\Delta t=\int_{\infty }^{R}\mathrm{d}R^{\prime }(R_{,t}-HR^{\prime }).
\end{equation*}%
This coincides (to leading order) with the expression for the virialised
case, eqn. (\ref{deltat}), for $R_{s}=0$. It seems natural, in the interior,
to consider the time derivative of $\phi $ w.r.t. $t^{\prime }$. As noted in
the previous section, $t^{\prime }$ will look like the Schwarzschild time
coordinate near the surface of a massive body (that is far outside its out
Schwarzschild radius), and $t^{\prime }\rightarrow t$ as $R\rightarrow
\infty $. We find that: 
\begin{equation*}
\phi _{,t^{\prime }}(r,t^{\prime })-\phi _{c,t}\sim -B_{,\phi }(\phi
_{c})\int_{\infty }^{R}\mathrm{d}R^{\prime }\Delta (R_{,t}\kappa \varepsilon
)(R^{\prime },t)+\ddot{\phi}_{c}(t)\Delta t+\mathcal{O}(\delta
_{1}^{2},\delta _{2}^{2},\delta _{3}^{3})
\end{equation*}%
where $\Delta (R_{,t}\kappa \varepsilon )(R,t)=R_{,t}\kappa \varepsilon
(R,t)-HR\kappa \varepsilon _{c}$. We note that this is the same as the
spherically-symmetric limit of the result found in eqn. (\ref{deltaphiv})
for the virialised case.

This completes the extension of our analysis to the case of spherically
symmetric collapsing spacetimes. It is clear that the quantity $\mathcal{I}$%
, defined in the analysis of the virialised case, will be also be a good
measure of $|(\dot{\phi}-\dot{\phi}_{c})/\dot{\phi}_{c}|$ in the collapsing
case. The validity of the matching procedure is this case is assured by the
condition that $\delta _{1},\delta _{2}\ll 1$ holds everywhere.

\section{Results and Consequences}

We now consider the astronomical consequences of our results for
observations here on Earth, and answer the basic question of whether local
experiments will detect cosmologically varying constants. We can evaluate
the quantity $\mathcal{I}$ explicitly for an Earth-based experiment assuming
the varying constant to be the Newtonian gravitation \textquotedblleft
constant" $G$ governed by Brans-Dicke theory (since in this case the
cosmological evolution of $\phi $ is easy to solve). We expect similar
values for BSBM, BM and other non-potential-dominated theories for varying $%
\alpha $ and $\mu $ \cite{bsm, bm}. 

We will consider a star (and associated planetary system) inside a galaxy
that is itself embedded in a large galactic cluster. The cluster is assumed
to have virialised and be of size $R_{clust}$. Close to the edge of the
cluster we allow for some dust to be unvirialised and still undergoing
collapse. There are three main contributions to $\mathcal{I}$ coming from
the star, the galaxy, and the galaxy cluster, respectively, and of these the
galaxy cluster contribution is by far the biggest. this can be understood by
noting that the galaxy cluster is the deepest gravitational potential well,
and the galaxy and star are only small perturbations to it. The contribution
to $\mathcal{I}$ from the galaxy cluster is found to be: 
\begin{eqnarray*}
\mathcal{I}_{clust} &\lesssim &\tfrac{3}{2}H_{0}(s-1/2)^{-1}\sqrt{%
2M_{clust}R_{clust}}\frac{\varepsilon _{clust}}{\varepsilon _{c}} \\
&=&\frac{10[3/(2s-1)]v_{clust}^{2}(1+z_{vir})^{3/2}\Delta _{vir}^{1/2}}{%
3\Omega _{m}^{1/2}}=1.01\times 10^{-5}\left( \frac{%
[3/(2s-1)](1+z_{vir})^{5/2}\Delta _{vir}^{5/6}}{\Omega _{m}^{1/6}}\right)
\left( \frac{hM_{clust}}{10^{15}M_{\odot }}\right) ^{2/3}, \\
&=&4.95\times 10^{-4}[3/(2s-1)]\Omega _{m}^{-1/2}\left( \frac{v_{clust}}{%
10^{3}\mathrm{km\,s^{-1}}}\right) ^{2}(1+z_{vir})^{3/2}, \\
&\approx &1.61\times 10^{-3}[3/(2s-1)]\Omega _{m}^{-1/2}(1+z_{vir})^{3/2}\ll
1
\end{eqnarray*}%
\noindent where we have used $3M_{clust}/5R_{clust}=v_{clust}^{2}=3\sigma
_{v}^{2}$ and $\kappa \varepsilon _{clust}=6M_{clust}/R_{clust}^{3}$; $%
\sigma _{v}$ is the 1-D velocity dispersion and $\Delta _{vir}\approx 178$
is the density contrast between the cluster and the background at
virialisation. In the final line of the approximation we have used the
representative value $\sigma _{v}=1040\mathrm{km\,s^{-1}}\Rightarrow
v_{vir}=1800\mathrm{km\,s^{-1}}$ appropriate for a rich cluster like Coma, 
\cite{peebles}. Taking a cosmological density parameter equal to $\Omega
_{m}=0.27,$ in accordance with WMAP, we expect that for a typical cluster
which virialised at a redshift $z_{vir}\ll 1$, we would have $\mathcal{I}%
_{clust}\approx 0.31[3/(2s-1)]\times 10^{-2}$. The term in $[..]$ is unity
when $s=2$, i.e. $2GM/R\rightarrow const$; such a matter distribution is
characteristic of dark matter halos. Different choices of $s>1/2$ only
change this estimate by a factor that is $\mathcal{O}(1)$. We note that,
since $2G\Delta M/R$ (where $\Delta M=M-\frac{1}{6}\varepsilon _{c}R^{3}$)
is required to be small as $R\rightarrow \infty $ by the matching
conditions, the model used here is only valid for $s\leq 2$ and hence the
singularity in $\mathcal{I}_{clust}$ at $s=1/2$ is fictitious. If we were to
have $\mathcal{I}_{clust}\gtrsim 1$ then we would require a large virial
velocity: $v_{clust}\gtrsim 32,400[3/(2s-1)]^{-1/2}(1+z_{vir})^{-3/4}\mathrm{%
km\,s^{-1}}$.

It is clear that in theories like Brans-Dicke, which have their cosmological
evolution dominated by the matter-to-dilaton coupling, $B_{,\phi }$, the
local time variation of $\phi $ and the associated constant differs from its
cosmological value by at most about $1\%$. In theories where the potential
dominates the cosmological evolution this result becomes even stronger and
we expect any deviations to occur only at the $0.4|B_{,\phi }(\phi
_{c})/V_{,\phi }(\phi _{c})|\%$ level, where $|B_{,\phi }(\phi
_{c})/V_{,\phi }(\phi _{c})|\ll 1$.

We have also seen that $\mathcal{I}$ is also good measure of $|(\dot{\phi}-%
\dot{\phi}_{c})/\dot{\phi}_{c}|$ inside spherically-symmetric regions that
are still undergoing gravitational collapse. Let us now evaluate $\mathcal{I}
$ for the case of a collapsing cluster in Brans-Dicke theory in the matter
era. We assume that the cluster is approximately homogeneous. We further
assume that when the cluster eventually virialises, at time $t_{vir}$, it
has a virialisation velocity $v_{vir}$. We use the spherical approximation
detailed in Chapter 5 of \cite{padmanabhan} to model the collapse of the the
cluster. We define $\theta $ by $t=t_{vir}(\theta -\sin \theta )/2\pi $, and
find: 
\begin{equation*}
\mathcal{I}(t)=\left\vert \frac{5v_{vir}^{2}}{2}\left( -\frac{\Delta
_{vir}^{1/2}f(\theta )(\theta -\sin \theta )}{2\sqrt{2}\pi }+\frac{\sqrt{2}%
\pi g(\theta )}{\Delta _{vir}^{1/2}(\theta -\sin \theta )}\right)
\right\vert =\left\vert \frac{5v_{vir}^{2}}{4}\left( -3f(\theta )(\theta
-\sin \theta )+\frac{2g(\theta )}{3(\theta -\sin \theta )}\right)
\right\vert ,
\end{equation*}%
\noindent where $\Delta _{vir}=18\pi ^{2}\approx 178$ is the density
contrast at virialisation when $\theta =2\pi $. When $\theta \leq 3\pi /2$ $%
f(\theta )=\sin \theta (1-\cos \theta )^{-3}$ and $g(\theta )=\sin \theta $;
for $\theta \geq 3\pi /2$, $f(\theta )=g(\theta )=-1$. In this evaluation we
have included the effect of the `drag term'; this is important up to
turnaround but it becomes negligible soon afterwards. 
\begin{figure}[tbh]
\begin{center}
\includegraphics[scale=0.85]{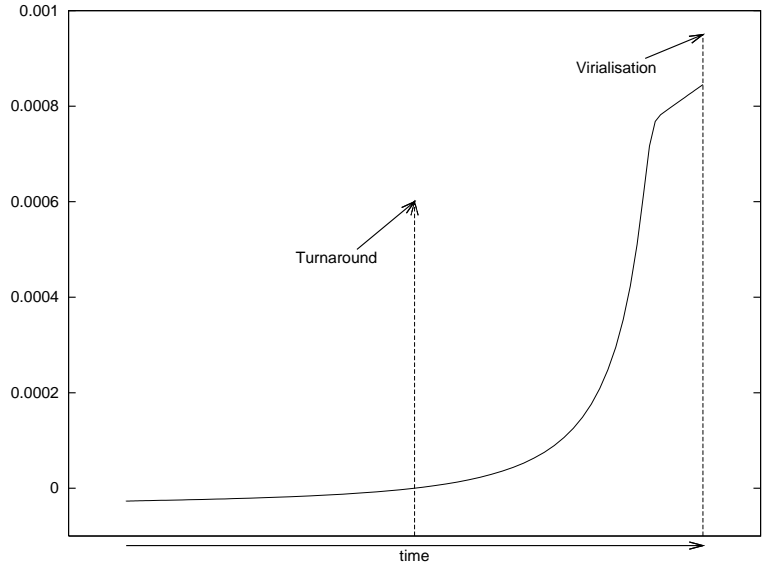}
\end{center}
\caption{Plot of $(\protect\phi _{,t^{\prime }}-\protect\phi _{c,t})/\protect%
\phi _{c,t}$ vs. time for Brans-Dicke theory at the centre of a collapsing
cluster with $v_{vir}=1800\mathrm{km\,s^{-1}}$.}
\label{fig1}
\end{figure}
Turnaround occurs at $\theta =\pi $, $t=t_{turn}$. In deriving the above
expression we have used $3M_{clust}/5R_{vir}=v_{vir}^{2}$, where $R_{vir}$
is the radius of the cluster after virialisation and $M_{clust}$ its mass.
The conditions required for the matching procedure to be valid are
equivalent to $10v_{vir}^{2}/(1-\cos \theta )\ll 1$, and it is clear that
this will not be satisfied all the way down to $\theta =0$. For $v_{vir}=1800%
\mathrm{kms^{-1},}$ our method will be valid for $\theta >0.027$ and for the
matching conditions to hold from turnaround to virialisation we require $%
v_{vir}\ll 95000\mathrm{km\,s^{-1}}$. Assuming that the cluster virialises
at an epoch that is close to the present day, this bound on $v_{vir}$
translates to requiring $R_{vir}\ll 432h^{-1}(1+z_{vir})^{-3/2}\mathrm{Mpc}$%
, where $H_{0}=100h\,\mathrm{km\,s^{-1}}\,\mathrm{Mpc}^{-1}$. We observe
that $\mathcal{I}$ is small up until turnaround and then grows quickly until
virialisation. At turnaround $\mathcal{I}=0,$ and at virialisation we find 
\begin{equation*}
\mathcal{I}(t_{vir})=\left( \frac{15\pi }{2}-\frac{5}{12\pi }\right)
v_{vir}^{2}=2.61\times 10^{-4}\left( \frac{v_{vir}}{10^{3}\mathrm{km\,s^{-1}}%
}\right) ^{2}\approx 0.85\times 10^{-3}.
\end{equation*}%
For the final evaluation we have taken $v_{vir}=1800\mathrm{km\,s^{-1}}$ (as
appropriate for the Coma cluster). The vanishing of $\mathcal{I}$ at
turnaround is specific to Brans-Dicke theory, more generally: $\mathcal{I}%
(t_{turn})=40v_{vir}^{2}[\ddot{\phi}_{c}-B_{,\phi }(\phi _{c})\kappa
\epsilon _{c}/H\dot{\phi}_{c}]/27\pi ^{2}$. In theories where the matter
coupling is \emph{strongly} dominant cosmologically, $|B_{,\phi }^{2}(\phi
_{c})\kappa \epsilon _{c}|\gg |V_{,\phi }(\phi _{c})|$, we find $\mathcal{I}%
(t_{turn})\approx 160v_{vir}^{2}|B_{,\phi \phi }(\phi _{c})|/27\pi ^{2}\ll 1$%
.

Our results differ greatly from those that were found using the \emph{%
spherical collapse model} used in \cite{bmot}; where $\mathcal{I}%
(t_{vir})\approx 200$. In that model, the spatial derivatives of $\phi $
were assumed to be negligible and are neglected. However, this is can only
be a realistic approximation when the collapsing region is as large as the
cosmological horizon; for a cluster virialising today that would require $%
R_{vir}\gtrsim 5\mathrm{Gpc}$. Since our method will fail for $%
R_{vir}\gtrsim 432h^{-1}(1+z_{vir})^{-3/2}\mathrm{Mpc}$, there must be some
region of intermediate behavior, $500\mathrm{Mpc}\lesssim R_{vir}\lesssim 5%
\mathrm{Gpc}$, that is not described by either the spherical collapse model
or our present analysis. We derived these results for Brans-Dicke theory,
where $\phi \propto G^{-1}$, however we should expect similar numbers for
all varying-constant theories where the cosmological dilaton evolution is
dominated by its matter coupling, $B_{,\phi }\kappa \varepsilon _{c}$. In
potential-dominated theories, the above numbers will be reduced by a factor
of $|B_{,\phi }(\phi _{c})\kappa \varepsilon _{c}/V_{,\phi }(\phi _{c})|\ll 1
$. As in the post-virialisation case, potential domination of the
cosmological evolution strengthens the amount to which local experiments
will see cosmologically varying constants.

In conclusion: we have used the method of matched asymptotic expansions to
find a sufficient condition for the time-variation of a scalar field, and
any related varying physical `constants' whose variation is driven by such a
field, to track its cosmological evolution. We have extended our previous
analyses by allowing `local' also to  include being inside some
spherically-symmetric \emph{collapsing} region. We have also proposed a
generalisation of our earlier condition for local variations to follow
global cosmological variations. We conjecture that this new condition is
applicable to scenarios more general than those we have explicitly
considered here. We have seen that this sufficient condition is always
satisfied for typical distributions of matter, and we have provided a proof
of what was previously merely assumed: \emph{terrestrial} and \emph{solar
system} based observations can legitimately be used to constrain the \emph{%
cosmological} time variation of many supposed `constants' of Nature.

\begin{acknowledgments}
We thank Timothy Clifton, David Mota and Peter D'Eath for discussions. DS
acknowledges a PPARC studentship.
\end{acknowledgments}

\end{document}